# Evidence of Scale-Free Topology and Dynamics in Gene Regulatory Networks

Sandy Shaw
Fractal Genomics, LLC
San Francisco, CA 94118-2603
sandy@fractalgenomics.com

**Abstract**

Temporal gene expression data (Wen, et. al.) was analyzed using the recently introduced inverse modeling technique of Embedded Complex Logistic Maps (ECLM). Preliminary results indicate a scale-free structure in the gene regulatory network topology and the network's energy dissipation characteristics. General agreement was found with other recent studies using dynamical models and connectivity analysis of other biochemical networks. The highlights of our study are briefly outlined.

## 1. Introduction

Coupled logistic maps [1,2] have been used to successfully model the dynamics of a variety of collective phenomena present in high-dimensional systems such as partial synchronization, power-law connectivity, and on-off intermittency [13-15, 17]. The ability to model high-dimensional systems with such low dimensional models appears to be aided by the properties of synchronization and long-range order, which help to effectively lower the dimensionality of the system [3, 7]. Despite these successes, and the effective lessening of the "curse of dimensionality" [6] brought about by these effects, reconstruction of the dynamics of complex, high-dimensional systems is a difficult task requiring far more data than can generally be acquired by experiment [5].

Wavelet analysis has been used to model aspects of high-dimensional systems directly from time series data [4]. A kind of localized Fourier Analysis, wavelet analysis is particularly effective in modeling transient or "bursty" behavior of systems [8]. Such behavior is present in a number of complex networks and systems, such as Internet traffic, financial market data, heartbeat time series, EEG data, etc. [8-10].

The method of Embedded Complex Logistic Maps (ECLM) [22,29] is a hybrid analysis method that blends some of the concepts behind wavelet and coupled logistic map methods to deal with the inverse modeling problem. ECLM is based on models within the complex plane that are derived from complex logistic maps (Julia sets [12]). Using ECLM, models are represented by points placed at particular coordinates in the complex plane based on how iterations of these coordinates, or orbits [12], best model individual system components.

To determine the best models for each system component, point-models are scored in their "natural" topology, within and near the Mandelbrot Set [11]. These models of system components, represented by points on the complex plane, are then compared with each other in the same topology. This comparison is aided by metrics on the surface that cluster similarities between point-models [22, 34-36]. By creating individual models and performing global comparisons in this way, ECLM is able to give a picture of local element dynamics in a global framework [22,29].

In this paper, we will discuss results of the method of ECLM applied to a temporal gene expression study of Rat CNS development (Wen, et. al.) [16]. These results will be compared with current theory and findings toward validation of ECLM as well as elucidating some potentially new findings involving scale-free topology and dynamics in gene regulatory networks.

## 2. ECLM Analysis of a Genetic Network

The input data for this ECLM trial consisted of gene expression data from 112 genes collected at nine different time points over 25 days, stretching from the embryonic stage through birth and post-birth, during rat central nervous system (cervical spinal cord and hippocampal) development [16]. Each of the 9 time points (including 100 corresponding snapshots or perturbations [22]) were fitted to an ECLM point model with $|\mathbf{C}| > .95$, where $\mathbf{C}$ is the Pearson correlation coefficient (assuming such a model could be found). All the fitted point models were then compared to each other to see which ones had pairwise $|\mathbf{C}| > .95$ for the 9 (model) time points. This led to 69 genes with a least one link to another gene (pairwise $|\mathbf{C}| > .95$ with at least one other ECLM model). We call these models linking ECLM (LECLM) [22].

### 2.1 Connectivity Distribution

Scale-free networks have been an area of much study in recent years [21]. In scale-free networks, $\mathbf{P(k)}$, the likelihood that a randomly chosen node from the network has $\mathbf{k}$ direct interactions, decays as a power law that is free of a characteristic scale [20]. Some recent results have shown scale-free structure in cellular metabolic networks and protein interaction networks. Some very recent results have indicated scale-free structure in transcription regulatory networks (global RNA expression) [18] and in gene expression networks [19]. It has also been suggested that all gene expression follows Zipf's law [23]. From the perspective of dynamics, scale-free structure has recently been found to spontaneously appear in coupled logistic maps when linking was based on dynamical considerations [17] rather than the notion of *preferential attachment* [24] usually cited in the evolution of scale-free networks.

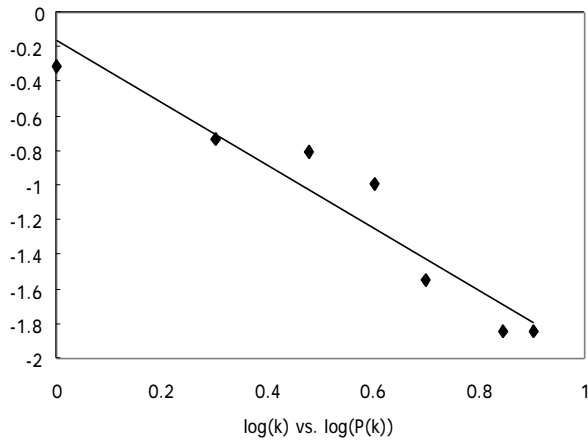

Fig. 1: Log-Log probability distribution for connectivity of linked ECLM models derived from genetic network data.

In Fig.1 you see the log-log plot of **P(k)** for the connectivity of LECLM models fitted to the temporal gene expression data from the Wen study. Although the data sample is small, the apparent scale-free structure in the data (linear correlation >.95) lends support to the assumptions of the dynamical model in the coupled logistic map study referenced above and appears to agree with the recent findings in other biochemical networks.

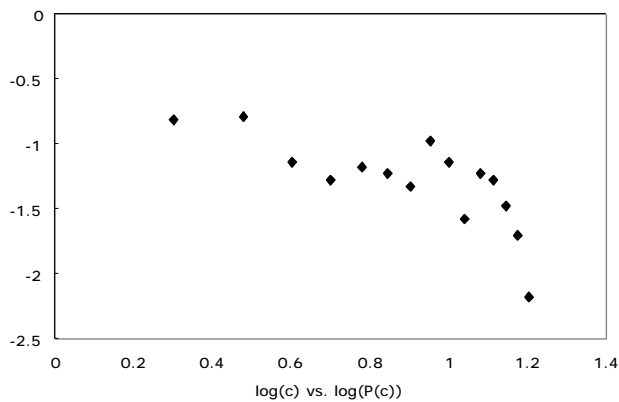

Fig. 2: Log-log probability distribution of **c** = cluster size.

The scaling constant of ~1.75 also falls in the range of scaling found in the metabolic and protein studies cited [25-28]. Besides lending support for the validity and usefulness of ECLM, this apparent scale-free connectivity is intriguing because it was produced from a dynamical perspective, using real data from a high-dimensional biological system, not just a model. The finding also has potential significance because the recent studies related to scale-free genetic networks were based on the organism S. cerevisiae, a yeast, and the Wen data is mammalian in origin.

### 2.2 Cluster Size Distribution

Fig. 2 is a log-log plot of the probability distribution of gene cluster size (cluster of genes whose behavior appeared correlated) for all LECLM. This distribution agrees quite well with the coupled logistic map study [17] where dynamic criteria was used to establish new links in the evolving network and scale-free structure was found. They found a cluster distribution with a power-law part followed by an exponential cutoff (which also begins near the end of this ECLM distribution in Fig. 2) after 100000 iterations of their model. In their study they also extended the cluster distribution curve to model systems with ~500 elements.

### 2.4 Energy Dissipation

Since this is a dissipative, open system, energy use and its relation to the system dynamics is an important consideration. One can interpret the number of LECLM as the number of ways available to dissipate energy into the system through "driving" or "driven" synchronization (correlated behavior) between genes. In this interpretation, if one rank orders the number of genes involved in one (correlated) gene cluster, two gene clusters, up to 8 (the maximum found for a single gene), then this can be viewed as an energy dissipation distribution for a randomly selected gene within the system. This is based on the notion that more energy will be dissipated into the system by a gene involved in a higher number of clusters or synchronization states [33]. Fig. 3 is a log-log plot of **P(M)**, the probability of a given gene driving **M** synchronization modes. Once again the data sample is small but a clear power-law scaling is seen (scaling constant ~2 and linear correlation > .96). A recent study suggested that almost all biological systems use hierarchical fractal-like networks as a way to minimize energy dissipation by efficiently transporting it between spatial scales [30] this effect occurs in other natural systems as well [31]. One can speculate that this finding is a variant of this type of energy dissipation minimization principle in a genetic network [32]. The interpretation of Fig. 3 as an energy dissipation PDF also seems to favorably match characteristics for other examples upon which the method of ECLM was applied [22].

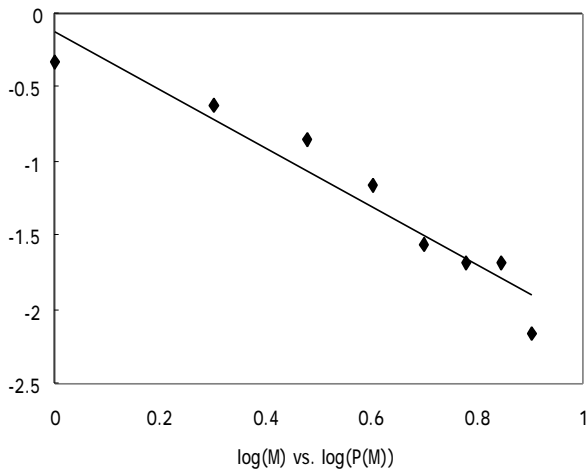

Fig. 3: Log-Log probability distribution of **M** = number of synchronization modes (energy states).

## 3. Discussion

This ECLM genetic network study produced a vast amount of information that is only just beginning to be analyzed [29, 32]. These preliminary results seem to indicate scale-free aspects in both the topology and dynamics of an actual gene regulatory network. General agreement with current studies also appears to be indicated. Diagrammatic analysis derived from ECLM (not shown in this paper) gives hints of an even deeper form of organization between "driving" and "driven" hubs in the network [29]. In light of recent evidence implying "universal" organizing principles in biological systems [37], one might speculate that these principles are at work on many levels, both structural and dynamic. With further studies and continuing validation of ECLM, we hope to extend these early findings and provide further evidence to support such speculation [32].

## References

[1] K. Kaneko, *Theory and Applications of coupled map lattices,* John Wiley and Sons Ltd., New York, 1993.

[2] K. Kaneko, ``Clustering, Coding, Switching, Hierarchical Ordering, and Control in Network of Chaotic Elements,'' *Physica D* **41**, 137-172, (1990).

[3] Y.-C. Lai, E. M. Bollt, and Z. Liu, ``Low-dimensional chaos in high-dimensional phase space: how does it occur?'' *Chaos, Solitons, and Fractals* **15**, 219-232 (2002).

[4] Patrice Abry, Richard Baraniuk, Patrick Flandrin, Rudolf Riedi, Darryl Veitch, "The Multiscale Nature of Network Traffic: Discovery. Analysis, and Modelling," *IEEE Signal Processing Magazine* **19**, no 3, pp 28-46 (May 2002).

[5] Y.-C. Lai and E. Kostelich, ``Detectability of dynamical coupling from delay-coordinate embedding of scalar time series,'' *Physical Review E* **66**, 036217(1-5) (2002).

[6] Bellman, R. *Adaptive Control Processes: A Guided Tour*, Princeton University Press (1961).

[7] Christos Faloutsos, Bernd-Uwe Pagel, and Flip Korn, "On the Dimensionality Curse and the Self-Similarity Blessing," *IEEE Transactions on Knowledge and Data Engineering* **13**, no 1, pp. 96-110 (Jan-Feb 2001).

[8] Shriram Sarvotham, Rudolf Riedi, Richard Baraniuk, "Connection-level Analysis and Modeling of Network Traffic," *Proceedings of the ACM SIGCOMM Internet Measurement Workshop, San Francisco, CA,* (Nov 2001).

[9] Goldberger AL, Amaral LAN, Glass L, Hausdorff JM, Ivanov PCh, Mark RG, Mietus JE, Moody GB, Peng CK, Stanley HE. PhysioBank, PhysioToolkit, and Physionet: Components of a New Research Resource for Complex Physiologic Signals. Circulation 101(23):e215-e220 [Circulation Electronic Pages; http://circ.ahajournals.org/cgi/content/full/101/23/e215]; 2000 (June 13).

[10] B. B. Mandelbrot. "Long-run linearity, locally Gaussian processes, H- spectra and infinite variances". *International Economic Review* **10**:82-113, 1969.

[11] Devaney, R. L. (ed.) *Complex Analytic Dynamics: The Mathematics Behind the Mandelbrot and Julia Sets*, Am. Math. Soc., Providence, 1994.

[12] Keen, L. Julia Sets. *In Chaos and Fractals: The Mathematics Behind the Computer Graphics*. Amer. Math. Soc., pp. 55-74, 1989.

[13] K. Kaneko, "Partition Complexity in Network of Chaotic Elements," *J. Phys*. A **24**, 2107-2119, 1991.

[14] A. Crisanti, M. Falcioni, and A. Vulpiani, "Broken ergodicity and glassy behavior in a deterministic chaotic map," *Phys. Rev. Lett*. **76**, 612-615, (1996).

[15] Mingzhou Ding and Weiming Yang, ``Stability of synchronous chaos and on-off intermittency in coupled map lattices'' *Physical Review E* **56**, no 4, pp. 4009-4016 (2002).

[16] Wen X., Fuhrman S., Michaels G.S., Carr D.B., Smith S., Barker J.L., Somogyi R., "Large-Scale Temporal Gene Expression Mapping of CNS Development". *Proc Natl Acad Sci* **95**:334-339, (1998).

[17] Pulin Gong and Cees van Leeuwen, "Emergence of scale-free network with chaotic units," *Sixth International Conference on Knowledge-Based Intelligent Information & Engineering Systems, Podere d'Ombriano, Crema, Italy* (Sept 2002).

[18] I.J. Farkas, H. Jeong, T. Vicsek, A.-L. Barabasi, Z.N. Oltvai, "The topology of the transcriptional regulatory network in the yeast, S. cerevisiae," Physica A **318,** 601, 2003.

[19] Featherstone, D. E., Broadie, K. "Wrestling with pleiotropy: genomic and topological analysis of the yeast gene expression network." *Bioessays* **24**:267-274, 2002.

[20] A.-L. Barabasi, and R. Albert, "Emergence of scaling in random networks," *Science* **286**, 509-512, 1999.

[21] R. Albert, A.-L. Barabasi, "Statistical mechanics of complex networks," *Rev. Mod. Phys*. **74**, 47 (2002).

[22] S. Shaw, "Inverse Modeling of Complex Networks Using Embedded Complex Logistic Maps," http://www.arxiv.org/abs/nlin.AO/0212049.

[23] K. Kaneko, "Zipf's Law in Gene Expression," http://www.arxiv.org/abs/physics/0209103.

[24] K. A. Eriksen, and M. Hornquist, "Scale Free Growing Networks Imply Linear Preferential Attachment," *Physical Review E* **65**, (2001):017102.

[25] Fell, D., and A. Wagner, "The small world of metabolism," *Nature Biotechnology* **18**: 1121-1122, 2000.

[26] Jeong, H., Tombor, B. Albert, R. Oltvai, Z.N., Barabasi, A.-L., "The large-scale organization of metabolic networks," *Nature* **407**: 651-654, 2000.

[27] Wagner, A., "The yeast protein interaction network evolves rapidly and contains few duplicate genes," *Molecular Biology and Evolution* **18**: 1283-1292, 2001.

[28] Wagner, A., and D. Fell, "The small world inside large metabolic networks," *Proc. Roy. Soc. London Ser. B* **268**, 1803-1810, 2000.

[29] S. Shaw, "Universal Modeling of Complex Networks,"http://www.fractalgenomics.com/papers/whitepaper_v10.pdf

[30] West GB, Woodruff WH, Brown JH, "Allometric scaling of metabolic rate from molecules and mitochondria to cells and mammals," *Proc Natl Acad Sci* **99**, Suppl 1:2473-8, (1998).

[31] Molnar, P.and J.A. Ramirez, "Energy dissipation theories and optimal channel characteristics of river networks," *Water Resour. Res*. **34**, 1809-1818, 1998.

[32] We will be exploring this question and other aspects related to life science in future papers and research. A series of technical reports are planned for on-line and/or paper publication. See http://www.fractalgenomics.com for more details. (Note: the web site is in the process of being reconstructed to reflect these latest results).

[33] O. Popovych, Yu. Maistrenko, E. Mosekilde, A. Pikovsky, J. Kurths, "Transcritical loss of synchronization in coupled chaotic systems," *Physics Letters A* **275**, 2000, 401–406.

[34] S. Shaw and C. Anthony Hunt, "A Novel Method for Visualizing Similarity in Gene Expression and Other Large Datasets," in *Proceedings of the ISCA 16th International Conference*, pp. 338-341, 2001.

[35] S. Shaw, "High Dimensional Mapping (HDM) Technology: A New Approach to Deal with Complex Data," Seminar at The Hubbard Center for Genome Studies, University of New Hampshire, Nov. 5, 2001.

[36] S. Shaw, "Fractal Formed Data Compression," NASA SBIR proposal No. 88-1-07-02-1468, 1988.

[37] Zoltán N.Oltvai and Albert-László Barabási, "Life's Complexity Pyramid". *Science* **298**, 763-764, 2002.